\documentclass[a4paper,11pt]{article}
\pdfoutput=1
\usepackage{jcappub}
\usepackage{subfigure}
\usepackage{epsfig}
\usepackage{graphicx}
\usepackage{color}
\usepackage{amssymb}
\usepackage{amsmath}
\usepackage{mathtools}
\usepackage{url}
\usepackage{natbib}
\usepackage[normalem]{ulem}
\usepackage{afterpage}
\usepackage{float}

\newcommand{\de}{\delta}
\newcommand{\De}{\Delta}

\newcommand{\Ga}{\Gamma}
\newcommand{\ka}{\kappa}

\newcommand{\La}{\Lambda}
\newcommand{\la}{\lambda}
\newcommand{\Om}{\Omega}

\newcommand{\be}{\begin{equation}}
\newcommand{\ee}{\end{equation}}
\newcommand{\gsim}{\stackrel{>}{\sim}}

\newcommand{\bea}{\begin{eqnarray}}
\newcommand{\eea}{\end{eqnarray}}
\newcommand{\bean}{\begin{eqnarray*}}
\newcommand{\eean}{\end{eqnarray*}}

\newcommand\spart{\;\raise1.0pt\hbox{$/$}\hskip-6pt\partial}
\definecolor{dgreen}{rgb}{0.0,0.6,0.1}

\title{The observable $E_g$ statistics}

\author[]{Basundhara Ghosh and}
\author[]{Ruth Durrer}
\affiliation[]{D\'epartement de Physique Th\'eorique and Center for Astroparticle Physics,\\
Universit\'e de Gen\`eve, 24 quai Ernest  Ansermet, 1211 Gen\`eve 4, Switzerland}

\emailAdd{basundhara.ghosh@unige.ch}
\emailAdd{ruth.durrer@unige.ch}

\abstract{Recently Moradinezhad Dizgah \& Durrer have shown that the $E_g$ statistics, useful to test theories of modified gravity, is plagued by additional scale and bias dependent lensing contributions. In this work we develop and illustrate a method to remove these lensing terms by using in addition to the galaxy clustering data also shear data and the correlations of shear and galaxy clustering. 
We introduce a truly observable statistics termed $\tilde E_g$ which conserves the properties of scale and bias independence on linear scales.
The method discussed here is best adapted to photometric surveys. It is found that the corrections to the original $E_g$ statistics are small for the present DES data, but for future surveys of the quality of Euclid they are very substantial.}


\arxivnumber{}
\begin{document}
\maketitle
\flushbottom

%-------------------------------------------------------------------------------
\section{Introduction}\label{s:intro}
Over the past few decades, our understanding of the observable Universe has undergone a rapid development, and especially observations of the cosmic microwave background (CMB) have led to a cosmological standard model, $\La$CDM, with parameters which are determined at a few percent precision or better. Lately, the inclusion of observations from the cosmological large scale structure (LSS), especially the baryon acoustic oscillations, has become more and more relevant. The observed accelerated expansion of the Universe is compatible with General Relativity (GR) only when including a cosmological constant $\Lambda$. 
Such a constant cannot be distinguished by any experiment from the effect of quantum vacuum energy, but its value, corresponding to an energy scale of about $10^{-3}$eV, is very discrepant with all expectations from particle physics.   This unsatisfactory situation has motivated the development of 
 theories of modified gravity, which challenge GR on very large scales, allowing for accelerated expansion without a cosmological constant. Thus, even though GR and modified gravity provide similar expansion laws at the background level,  they usually exhibit differences when the evolution of perturbations are taken into account. This is something that can be effectively probed with current and upcoming cosmological LSS surveys.

A recently proposed observable which aims to distinguish between $\Lambda$CDM and modified gravity theories by providing a test for gravity at large scales, is the so-called $E_g$ statistics introduced by Zhang et al. in 2007~\cite{Zhang:2007nk}. The $E_g$ statistics is defined as the ratio
of galaxy-lensing cross-correlations to the galaxy-velocity cross-correlations, and is expected to provide a bias independent insight if both measurements are made using the same galaxy population. There have been different measurements of $E_g$ using different data sets since then~\cite{Reyes:2010tr,Blake:2015vea,Pullen:2015vtb}. In this paper, we will follow the approach used by Pullen et al.~\cite{Pullen:2015vtb},  where instead of the galaxy-lensing cross correlations, the CMB lensing cross correlations are used. The reason for this is that CMB lensing is very accurate and enables surveys to probe at earlier times and even larger scales compared to galaxy surveys. Also, it is not plagued by the systematics of shear measurements  such as intrinsic alignment.

A previous study~\cite{Dizgah:2016bgm} has shown, however, that neglecting the lensing contribution to galaxy number counts at high redshifts can give rise to large errors (up to 
25-40 \% for $z=1.5$), and is hence indispensable for correct number count measurements. However, if the definition of $E_g$ includes these corrections, the scale and bias independence of the $E_g$ statistics is compromised, which refutes its original purpose. As a result, one needs to develop a method which will enable the desired use of this quantity without posing a threat to its special, positive features. 

In this paper, we explore how lensing corrections can be incorporated in standard $E_g$ measurements. However, while~\cite{Dizgah:2016bgm} suggest to use 21 cm intensity mapping which is not affected by lensing and which has been explored in~\cite{Pourtsidou:2015ksn}, we  propose a different method which can be implemented in galaxy number count surveys. Although intensity mapping measurements have a promising future with surveys like the Square Kilometer Array (SKA)~\cite{Maartens:2015mra} , we should also be able to carry out studies regarding $E_g$ statistics with the ongoing galaxy surveys.
 
 A correction similar to the one proposed in the present work has also been suggested in Ref.~\cite{Yang:2018boi}. In this paper, however the lensing contribution was estimated by rescaling the CMB lensing term. Here we propose to use in addition to the number counts also shear data which directly measure
the lensing power spectrum and the galaxy--lensing cross correlation spectrum at the redshift of the galaxy survey. This avoids the uncertain, and most probably bias dependent scaling proposed in~\cite{Yang:2018boi}, but requests more observational data. For our method to work we need therefore both, number count and shear power spectra at a given redshift $z$ as well as the number count--CMB lensing correlation spectrum and the CMB lensing power spectrum.

According to~\cite{Pullen:2015vtb}, $E_g$ in case of CMB lensing can be estimated as,
\bea
E_g(\ell,z)=\Gamma(z)\frac{C_\ell^{\ka g}(z_*,z)}{\beta C_\ell^{gg}(z,z)}
\label{e:egdef}
\eea
where $z_*$ denotes the CMB redshift and $z$ is the redshift of the galaxy survey. 
$\Gamma(z)$ is a pre-factor depending on Hubble parameter
$H(z)$, the  lensing kernel $W(z_*,z)$, and the galaxy redshift distribution denoted by $f_g(z)$. The pre-factor $\Gamma(z)$ is not very relevant for this work as we simply wish to provide a method of measuring $E_g$ without going into the technicalities, and hence we do not discuss it in detail here. More about this pre-factor can be found in literature~\cite{Pullen:2015vtb,Yang:2018boi}.  $C_\ell^{\ka g}$ is the lensing convergence--galaxy angular cross-power spectrum, 
$C_\ell ^{gg}$
is the galaxy angular auto-power spectrum, and $\beta$ is the redshift space distortion parameter. Within linear perturbation theory, $\beta=f/b_g$, where $f$ is the linear growth rate and $b_g$ is the galaxy bias.

The remainder of this paper is structured as follows: In Section~\ref{s:theo} we briefly outline the theory leading to $E_g$, and explain how lensing corrections can be effectively incorporated in our measurements. In Section~\ref{s:num} we present numerical results for CMB lensing of foreground galaxies with redshift bins corresponding to the Dark Energy Survey (DES) Year 1 results~\cite{Abbott:2017wau} and for some redshifts of the Euclid collaboration~\cite{Amendola:2016saw}. Our numerical results are obtained from linear perturbation theory and are therefore only valid on sufficiently large scales. In Section~\ref{s:con} we discuss our findings and conclude.

\section{Theory}\label{s:theo}
As mentioned above, $E_g$ statistics is aimed at  distinguishing between GR and modified gravity theories. The fundamental parameters that come to mind when we wish to make this distinction are the gravitational Bardeen potentials $\Phi$ and $\Psi$ which appear in the perturbed Friedman-Robertson-Walker metric in longitudinal gauge,
\bea
ds^2=a^2(t)[-(1+2\Psi)dt^2+(1-2\Phi)\delta_{ij}dx^idx^j] \,.
\eea
In the case of GR, $\Phi=\Psi$ in the absence of anisotropic stress. However, for theories of modified gravity, a valid way to check the relation between these two potentials is to measure the ratio between the lensing effect that is proportional to the quantity $\nabla^2(\Phi+\Psi)$ and the peculiar velocity field of non-relativistic particles that is related to the time component of the metric,  $\Psi$. Following Zhang~\cite{Zhang:2007nk}, we define in  Fourier space,
\bea
E_g(z,k)=\frac{k^2(\Phi+\Psi)}{3H_0^2(1+z)\theta(k)}\,,
\label{e:zhang}
\eea 
where $\theta=\nabla\cdot\textbf{v}/H(z)$, \textbf{v} being the peculiar velocity field and $H(z)$ the Hubble parameter at a redshift $z$. The quantity $\theta(k)$ in $\Lambda$CDM cosmology becomes $\theta(k)=f(z)\delta_m(k,z)$, where $f(z)\simeq[\Omega_m(z)]^{0.55}$ within GR.  $\Om_m(z)$ denotes the matter density parameter at redshift $z$. According to Poisson's equation,
\be
k^2\Phi=\frac{3}{2}H_0^2\Omega_{m,0}(1+z)\delta_m\,,
\label{e:poisson}
\ee
where $\Om_{m,0}=\Om_m(0)$.
Thus, within GR, $E_g$ reduces to $E_g(z,k)=\Omega_{m,0}/f(z)$, simplifying eq.(\ref{e:zhang}) with the help of eq.(\ref{e:poisson}) and the relation $\Phi=\Psi$. In particular, within linear perturbation theory $E_g$ depends neither on scale nor on galaxy bias.

This reduced version of $E_g$ can be given a more general form in order to cater to modified gravity by incorporating two arbitrary functions $\mu(k,z)$ and $\gamma(k,z)$ such that
\be
k^2\Phi=4\pi G a ^2\bar{\rho}\mu(k,z)\delta_m(k,z),\quad
\Psi=\gamma(k,z)\Phi \,.
\ee
In a theory where the modification of gravity can be cast in this way, we have,
\be
E_g(k,z)=\frac{\Omega_{m,0}\mu(k,z)[\gamma(k,z)+1]}{2f}
\ee
which reduces to the GR form for $\mu(k,z)=\gamma(k,z)=1$. For a generic theory of modified gravity, however, we expect this quantity to depend on scale in a non-trivial way. Of course also non-linearities of gravity will induce a scale dependence on smaller scales. We therefore have to be careful to use this statistics only on large enough scales.

Having discussed the basic idea behind the $E_g$ statistics approach, it is important to understand why lensing corrections are inevitable as well as a threat to the very foundation of $E_g$ statistics, and how we can deal with them. In Section~\ref{s:num}, we show some numerical examples to support our claim, but in this section, we wish to assert its importance theoretically. 

For our current interests, we will be focussing only on the over-density and lensing term (for the full expression, see~\cite{Bonvin:2011bg}). We can neglect the redshift space distortion for wide redshift bins having $\Delta z \gsim 0.1$, and  we may also ignore large-scale effects involving the Bardeen potentials for subhorizon scales ($\ell>20$). Thus we can approximate the galaxy fluctuations in direction \textbf{n} and at redshift z by:
\bea
\nonumber
\Delta^g(\textbf{n},z)&=&b(z)\delta_m(\chi(z)\textbf{n},z)+\left(1-\frac{5}{2}s(z)\right)\int_{0}^{\chi(z)}d\chi\frac{\chi(z)-\chi}{\chi(z)\chi}\nabla_\Omega^2(\Phi+\Psi)(\chi\textbf{n},t_0-\chi) \\
&=&b\delta_m - 2\left(1-\frac{5}{2}s\right)\kappa
\eea
where $\chi(z)$ is the comoving distance to redshift $z$, $b(z)$ is the galaxy bias and $s(z)$ is the magnification bias. $\delta_m$ is the matter over-density and $\kappa$ is the convergence.\\

The magnification bias is the logarithmic derivative of the galaxy number count at the limiting magnitude,
\be
s(z,m_{\rm lim}) \equiv \left.\frac{\partial\log_{10}{\bar N}(z,L> L_{\rm lim})}{\partial m}\right|_{m_{\rm lim}} \,.
\label{e:s_mlim}
\ee
Here $m_{\rm lim}= 5\log L_{\rm lim} +$const.  is the limiting magnitude of the survey. The magnification bias accounts for the fact that  highly magnified galaxies, even if they are intrinsically not luminous enough can get included in the survey. Only if the survey is sensitive enough to include all galaxies (of a given type) down to the lowest luminosities, we have $s=0$.

Since we truly observe $\De^g$ and not $\de_m$, if we have to take along the additional lensing term, $\ka$ , with the overdensity $\delta$ when correlating number counts with lensing data.  For any two redshifts $z_1$ and $z_2$ we then find
\be
C_\ell^{\kappa g}(z_1,z_2)=b(z_2)C_\ell^{\kappa\delta}(z_1,z_2)-(2-5s(z_2))C_\ell^{\kappa\kappa}(z_1,z_2)
\label{e:num}
\ee
and correspondingly, the auto-correlation of number counts gives
\bea
C_\ell^{gg}(z_1,z_2) &=&b(z_1)b(z_2)C_\ell^{\delta\delta}(z_1,z_2)+(2-5s(z_1))(2-5s(z_2))C_\ell^{\kappa\kappa}(z_1,z_2) \nonumber\\
&& -b(z_2)(2-5s(z_1))C_\ell^{\kappa\delta}(z_1,z_2)-b(z_1)(2-5s(z_2))C_\ell^{\kappa\delta}(z_2,z_1)
\label{e:deno}
\eea
Now, in case of CMB lensing, the scale independent and bias independent quantity actually is not the one given in eq.(\ref{e:egdef}) but
\bea
\tilde{E_g}(\ell,z)=\Gamma(z)\frac{C_\ell^{\ka\delta}(z_*,z)}{f(z) C_\ell^{\delta\delta}(z,z)} \,,
\label{e:deftilde}
\eea
while what we naively measure is
\bea
E_g(\ell,z)=\Gamma(z)\frac{C_\ell^{\ka g}(z_*,z)}{\beta(z) C_\ell^{gg}(z,z)} \,.
\label{e:def}
\eea
Here, $E_g$ is the statistics that galaxy surveys actually measure, while $\tilde{E_g}$ is the one excluding lensing corrections. Inserting (\ref{e:num}) and (\ref{e:deno}) above we have
\be
E_g\propto\frac{C_\ell^{\ka g}(z_*,z)}{\beta(z) C_\ell^{gg}(z,z)}=\frac{1}{\beta(z)}\frac{bC_\ell^{\kappa\delta}(z_*,z)-(2-5s)C_\ell^{\kappa\kappa}(z_*,z)}{b^2C_\ell^{\delta\delta}(z,z)+(2-5s)^2C_\ell^{\kappa\kappa}(z,z)-2b(2-5s)C_\ell^{\kappa\delta}(z,z)} \,,
\label{e:delkap}
\ee
where $b$ and $s$ are to be evaluated at redshift $z$ at all instances.
It is clear from eq.~(\ref{e:delkap}) that a straightforward way to remove the difference between the scale independent quantity $\tilde E_g$ and the measured quantity $E_g$ would be to have a galaxy population with $(2-5s)=0$, i.e., $s=2/5$ which corresponds to intensity mapping. This is the idea suggested in~\cite{Dizgah:2016bgm,Pourtsidou:2015ksn}, but here we want to explore a different method that can be implemented in galaxy surveys, without depending on  upcoming intensity mapping surveys. The idea is to obtain the $E_g$ statistics for the density power spectra in terms of the galaxy number counts, the lensing power spectrum (obtained by shear measurements) and their correlation, that is, in terms of the observable spectra, namely, $C_\ell^{\ka g}(z_1,z_2)$, $C_\ell^{gg}(z_1,z_2)$ and $C_\ell^{\ka\ka}(z_1,z_2)$ which appear in eq.~\eqref{e:num} and eq.~\eqref{e:deno}.

Using  (\ref{e:num}) we first find
\be
b(z)C_\ell^{\kappa\delta}(z_*,z) = C_\ell^{\ka g}(z_*,z) +(2-5s(z))C_\ell^{\kappa\kappa}(z_*,z)\,.
\ee
From~(\ref{e:deno})  we obtain
\bea
\hspace*{-0.5cm}b^2(z)C_\ell^{\de\delta}(z,z) &=&C_\ell^{g g}(z,z) -(2-5s(z))^2C_\ell^{\kappa\kappa}(z,z)+2b(z)(2-5s(z))C_\ell^{\kappa\delta}(z,z) \,, \nonumber \\
&=& C_\ell^{g g}(z,z) +(2-5s(z))^2C_\ell^{\kappa\kappa}(z,z)+2b(z)(2-5s(z))C_\ell^{\kappa g}(z,z) \,.
\eea
Inserting this in eq.~\eqref{e:deftilde}  we obtain
\be
\tilde{E_g}(\ell,z)=\Gamma(z)\frac{bC_\ell^{\ka\delta}(z_*,z)}{\beta(z) b^2 C_\ell^{\delta\delta}(z,z)}=\frac{1}{\beta(z)}\frac{C_\ell^{\kappa g}(z_*,z) + (2-5s)C_\ell^{\kappa\kappa}(z_*,z)}{C_\ell^{gg}(z,z)+(2-5s)^2C_\ell^{\kappa\kappa}(z,z)+2(2-5s)C_\ell^{\kappa g}(z,z)} \,,
\label{e:subtract}
\ee
where $b$ and $s$ are to be evaluated at redshift $z$. To measure $\tilde E_g$ we therefore need to measure not only $C_\ell^{\kappa g}(z_*,z)$ and $C_\ell^{gg}(z,z)$, but also $C_\ell^{\kappa\kappa}(z_*,z)$,  $C_\ell^{\kappa\kappa}(z,z)$ and $C_\ell^{\kappa g}(z,z)$ in order to correct for the lensing contamination in  $C_\ell^{\kappa g}(z_*,z)$ and  $C_\ell^{gg}(z,z)$.

In the next section we want to study in which cases  the difference between $\tilde E_g$ and  
$E_g$  is relevant. For this we introduce the relative difference defined as,
\be
\frac{\Delta E_g}{\tilde{E_g}}=\frac{\tilde{E_g}-E_g}{\tilde{E_g}}\,.
\label{e:rel}
\ee

\section{Numerical Results}\label{s:num}
In order to illustrate our method proposed in Section~\ref{s:theo}, we show some examples using the specifications from the Dark Energy Survey (DES) Year 1 and for a Euclid-like survey~\cite{EuclidRB,Amendola:2016saw}.
For the numerical calculation we use the public code {\sc class}~\cite{Lesgourgues:2011re,Blas:2011rf} in which relativistic contributions to galaxy number counts are included~\cite{DiDio:2013bqa}. We assume purely scalar perturbations, and consider the cosmological parameters of the Planck 2015 results~\cite{Planck:2015xua}. Thus in our evaluations, the Hubble parameter is $H_0=67.556$~km s$^{-1}$ Mpc$^{-1}=100h$~km s$^{-1}$ Mpc$^{-1}$, the baryon density parameter is $\Omega_bh^2=0.022032$, the cold dark matter density parameter is $\Omega_{\rm cdm}h^2=0.12038$, the curvature is $K=0$, the number of neutrino species is $N_\nu=3.046$ and the neutrino masses are neglected. The bias parameter $b(z)$ is taken from the literature as discussed below and the galaxy power spectrum is obtained from the matter power spectrum by multiplication with this linear bias factor as given in eqs.~(\ref{e:num}) and (\ref{e:deno}).

\subsection{$E_g$ statistics for DES-like redshift binning}\label{sub:des}
From the specifications of the first year results of the DES collaboration~\cite{Abbott:2017wau,Troxel:2017xyo,Drlica-Wagner:2017tkk,Hoyle:2017mee,Elvin-Poole:2017xsf}  which has measured all - the galaxy number counts, lensing and  the cross-correlation of galaxy clustering and lensing, we consider five galaxy redshift bins in the foreground having a width $\De z=0.15$.
The mean redshifts of these bins are,
\be
z_1= 0.225\quad z_2= 0.375\quad z_3=0.525 \quad z_4=0.675 \quad  z_5=0.825\,.
\label{e:desz}
\ee
In this section, we compute the required angular power spectra for the purpose of $E_g$ measurement, in case of CMB lensing for the above mentioned DES foreground redshifts using both linear perturbation theory and the halofit approximation for the matter power spectrum. For our purpose, we assume a complete survey setting\footnote{The DES collaboration has not published any value for $s(z)$, the true value is most probably different from zero and depends on $z$. However it is not available to us as the DES collaboration has neglected convergence in its analysis.} $s=0$ and galaxy bias parameters as given in fig.˜7 of~\cite{Abbott:2017wau}. We have  shown in a previous paper~\cite{Ghosh:2018nsm} that neglecting lensing effects in case of DES-like redshift binning can cause up to 50\% error in the density-shear correlations, and hence it is expected to cause a significant error in $E_g$ measurements as well.

\begin{figure}[H]
	%\vspace*{-1cm} ~~  \\
	\centering
	{\includegraphics[width=0.8\textwidth]{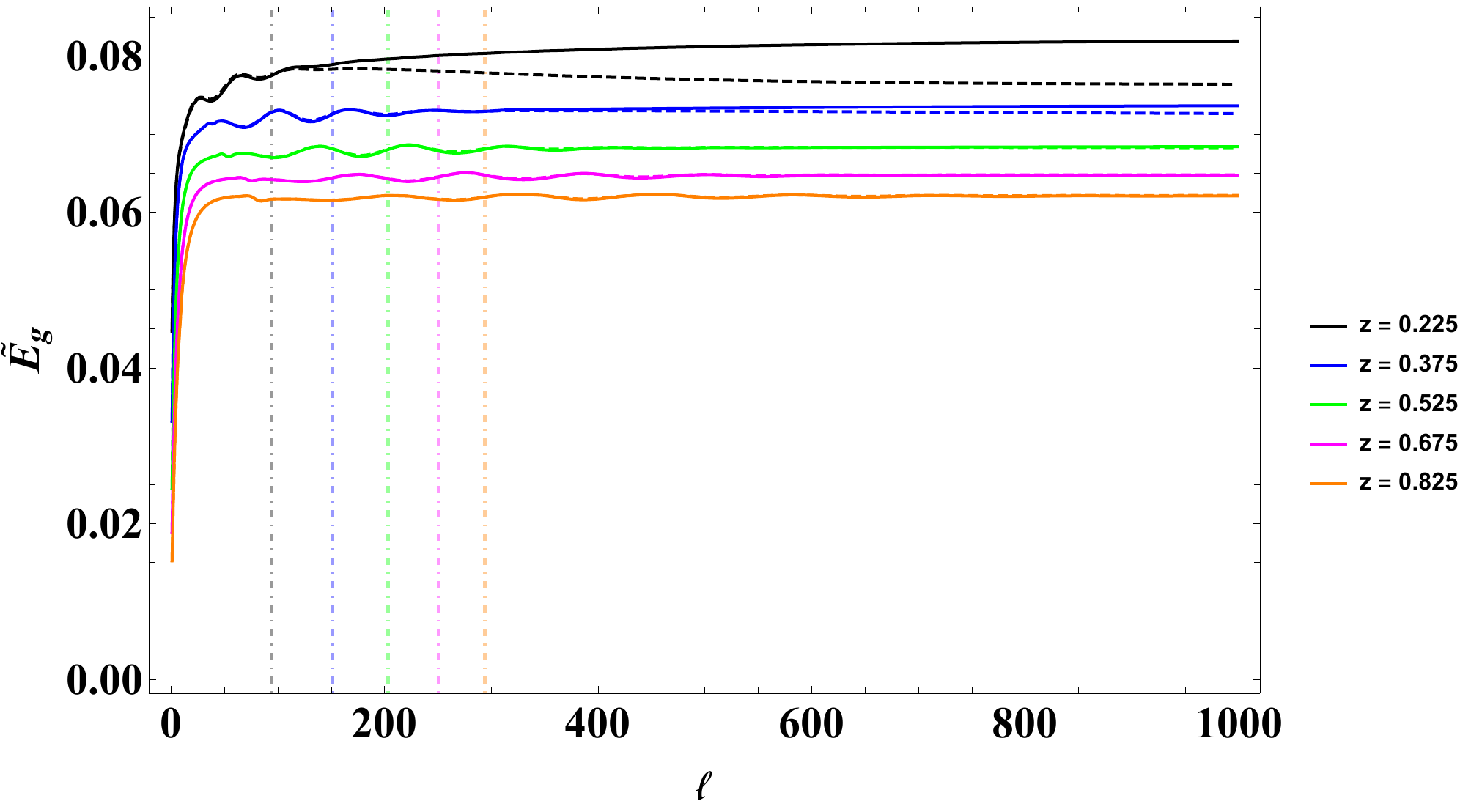}}\par
   \hspace*{0.17cm} {\includegraphics[width=0.815\textwidth]{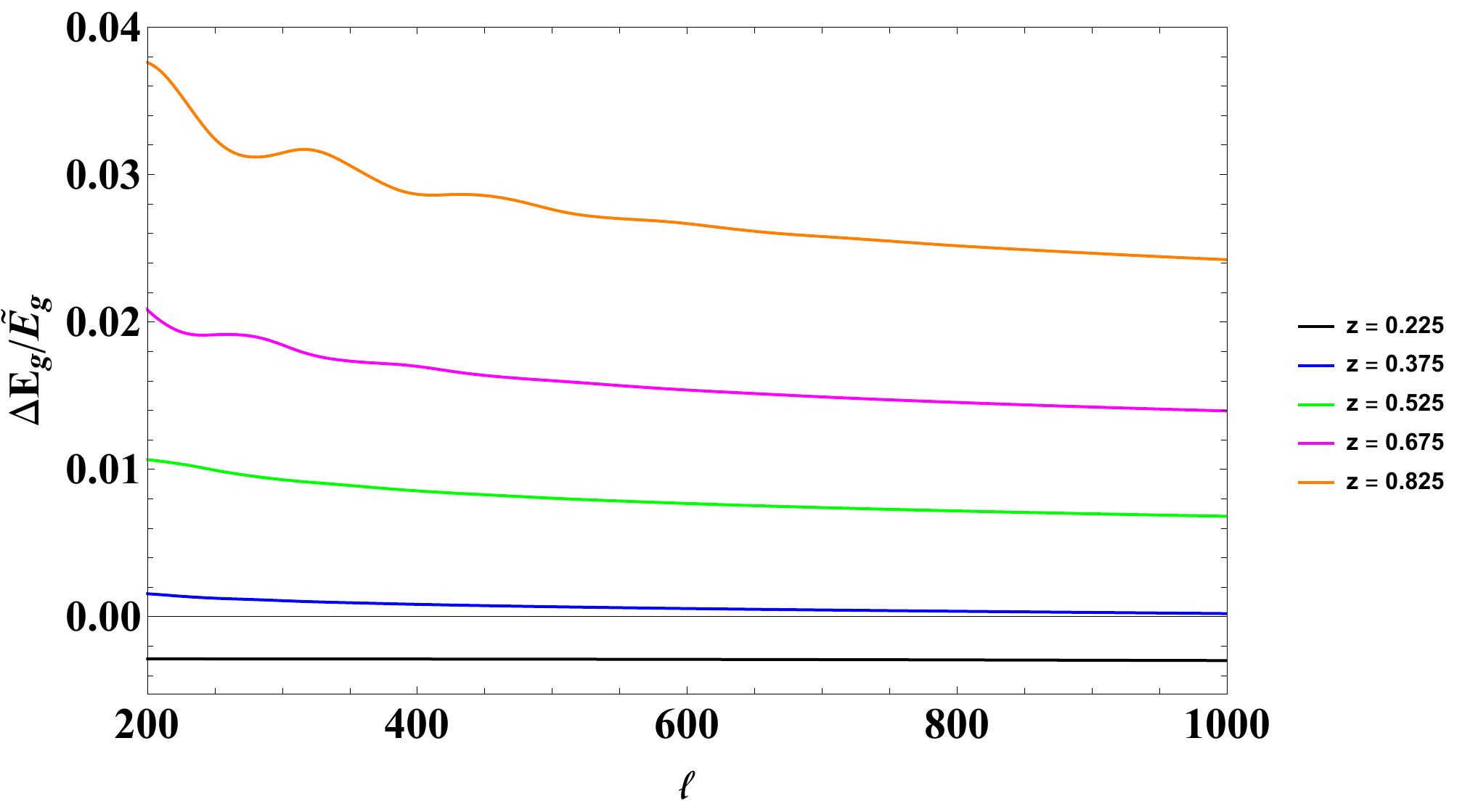}}
	\caption{The upper panel shows $\tilde{E}_g$ as a function of $\ell$ for different values of $z$ corresponding to foreground bins of DES Year 1 results. The solid lines are obtained using linear theory, and the dashed correspond to Halofit. The vertical dot-dashed lines signify the value of $\ell=\ell_L$ greater than which nonlinearities become prominent, where we have used $\ell_L=k_L \chi(\bar{z})$, with $k_L=0.1$ \textit{h}/Mpc  and $\chi(\bar{z})$ being the comoving distance for the mean redshift $\bar{z}$ of a particular redshift slice. The lower panel shows the correction on $E_g$ value due to lensing effects for each redshift bin.}
	\label{f:des}
\end{figure}

In fig.~\ref{f:des} we show the behaviour of $\tilde{E}_g(\ell,z)$ against $\ell$ for different values of DES foreground redshifts, which have been obtained from the observable angular power spectra using eq.~\eqref{e:subtract}. $\tilde E_g$ is clearly scale independent for $\ell\gsim 100$. We indicate as a vertical line the redshift dependent linearity scale,  $\ell_L=k_L \chi(\bar{z})$, for $k_L=0.1h/$Mpc  where $\chi(\bar{z})$ denotes the comoving distance for the mean redshift  of the bin. Note that $\la_L=2\pi/k_L$ is the wavelength corresponding to $\ell$ only for exactly equal redshifts. Since our redshift bins are relatively wide, the mean wavelength corresponding to $\ell_L$ will have a non-negligible component in the radial direction and will therefore be substantially larger. We also compute the uncorrected $E_g$ and  find that it has an error of about 4\% compared to the corrected one, $\tilde E_g$, for the highest foreground redshift of DES.  This error is much smaller than the ratio $C_\ell^{\ka\ka}/C_\ell^{\de\ka}$ which can become as large as 0.3, see Ref.~\cite{Ghosh:2018nsm}.  We suggest that the additional terms in the numerator and the denominator partially cancel in the correction. As we shall see in Section~\ref{sub:euclid}, for much higher redshifts like that of Euclid, this error   increases by about a factor of ten.

Interestingly, at low $\ell$ we see a strong scale dependence. This is due to the fact that the $k$-space scale invariance of $E_g(k,z)$ of Eq.~(\ref{e:zhang}) translates into an $\ell$-invariance of $E_g(\ell,z)$ given in Eq.~(\ref{e:def}) only in the flat sky limit, which at sufficiently high redshift is reached at $\ell\sim 30$, while at the lowest redshift it is valid only after $\ell\sim 100$ where corrections due to non-linearities already become important. Therefore, this lowest redshift bin is not very useful for the $E_g$ statistics. It is however very promising that for the higher redshifts $z\geq 0.52$ the difference between the halofit result (dashed line) and the linear perturbation theory result (solid line) is negligible. This is partially due to the fact that at higher redshift non-linearities are less important, but it mainly comes from the fact that we use relatively wide redshift bins for which the small scale structure is significantly damped. Therefore our $E_g$-statistics for these redshifts can be used up to $\ell\sim 1000$. The slight 'wiggles' seen in $E_g$ are the baryon acoustic oscillations which are seen in the galaxy distribution but are integrated over in $\kappa$. As expected they move to higher $\ell$'s at higher redshift and they are slightly damped by non-linearities.

\subsection{$E_g$ statistics for Euclid-like redshift binning}\label{sub:euclid}
Now we repeat the same analysis as in Section~\ref{sub:des} for a few redshifts probed by the Euclid satellite~\cite{Amendola:2016saw}. According to Euclid photometric specifications, the galaxy bias and magnification bias are given as a function of the redshift as\footnote{In Ref.~\cite{Yang:2018boi} Euclid is also considered with a similar galaxy bias, but the luminosity bias is set to the constant values $s=0.48$ for the spectroscopic survey and $s=0.326$ for the photometric survey. These values are close to the point $s=0.4$ where the correction vanishes, artificially reducing the lensing correction.},
\bea
b(z)=b_0\sqrt{1+z}\\
s(z)=s_0+s_1z+s_2z^2+s_3z^3
\eea
Here we set $b_0=1$ and the magnification bias coefficients are $s_0=0.1194$, $s_1 = 0.2122$, $s_2 =-0.0671$ and $s_3=0.1031$. The window function is taken to be Gaussian with a standard deviation of $\Delta z_i/2$, where $\Delta z_i\gsim 2\delta_z$ is the width of the $i$-th bin, $\delta_z$ being the photometric redshift error given by $\delta_z=0.05(1+z)$. For more details about Euclid specifications, we also refer to Appendices A.1 and B of~\cite{Montanari:2015rga}.

\begin{figure}[H]
	%\vspace*{-1cm} ~~  \\
	\centering
	{\includegraphics[width=0.8\textwidth]{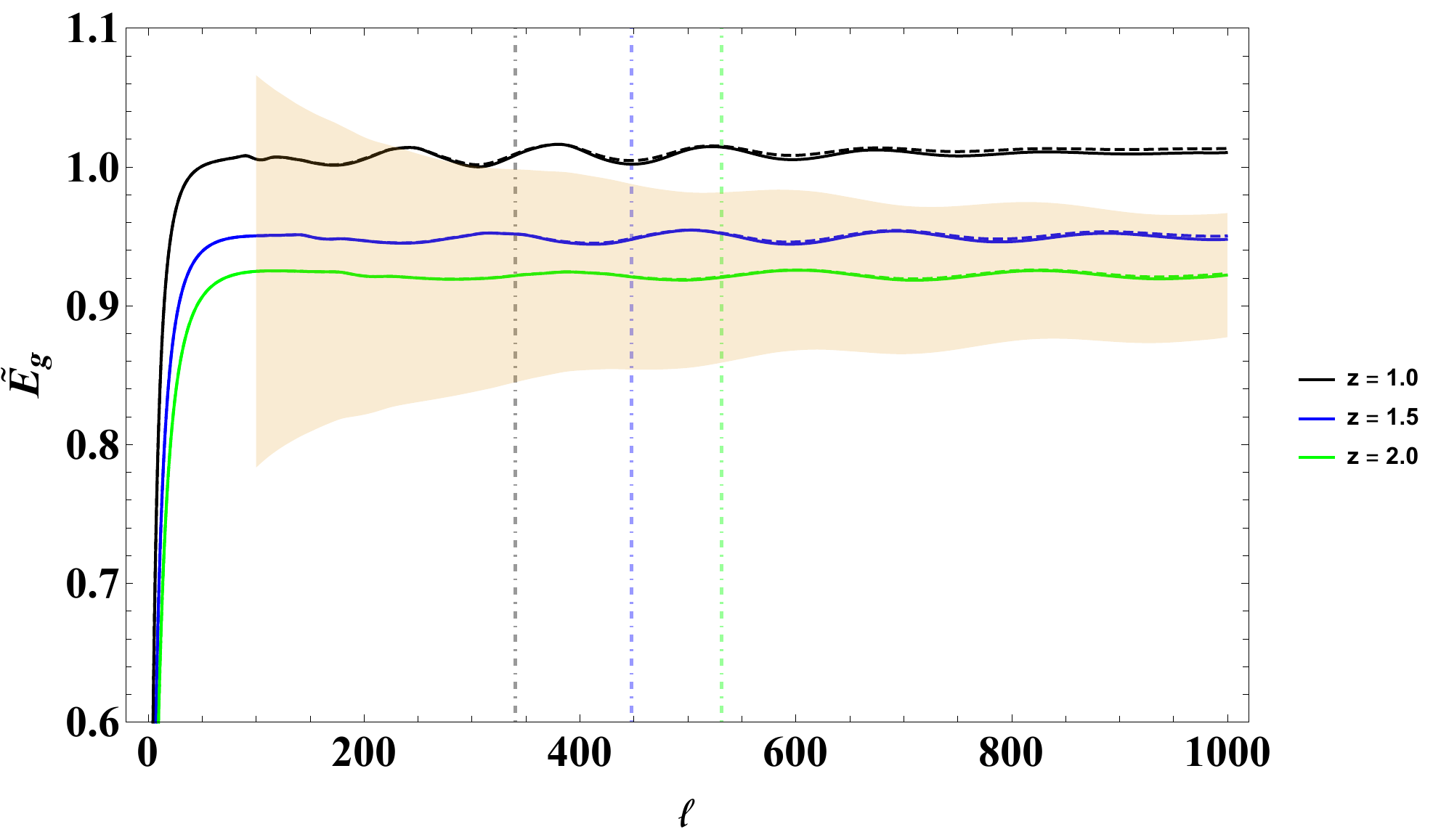}}\par
\hspace*{-0.05cm} {\includegraphics[width=0.84\textwidth]{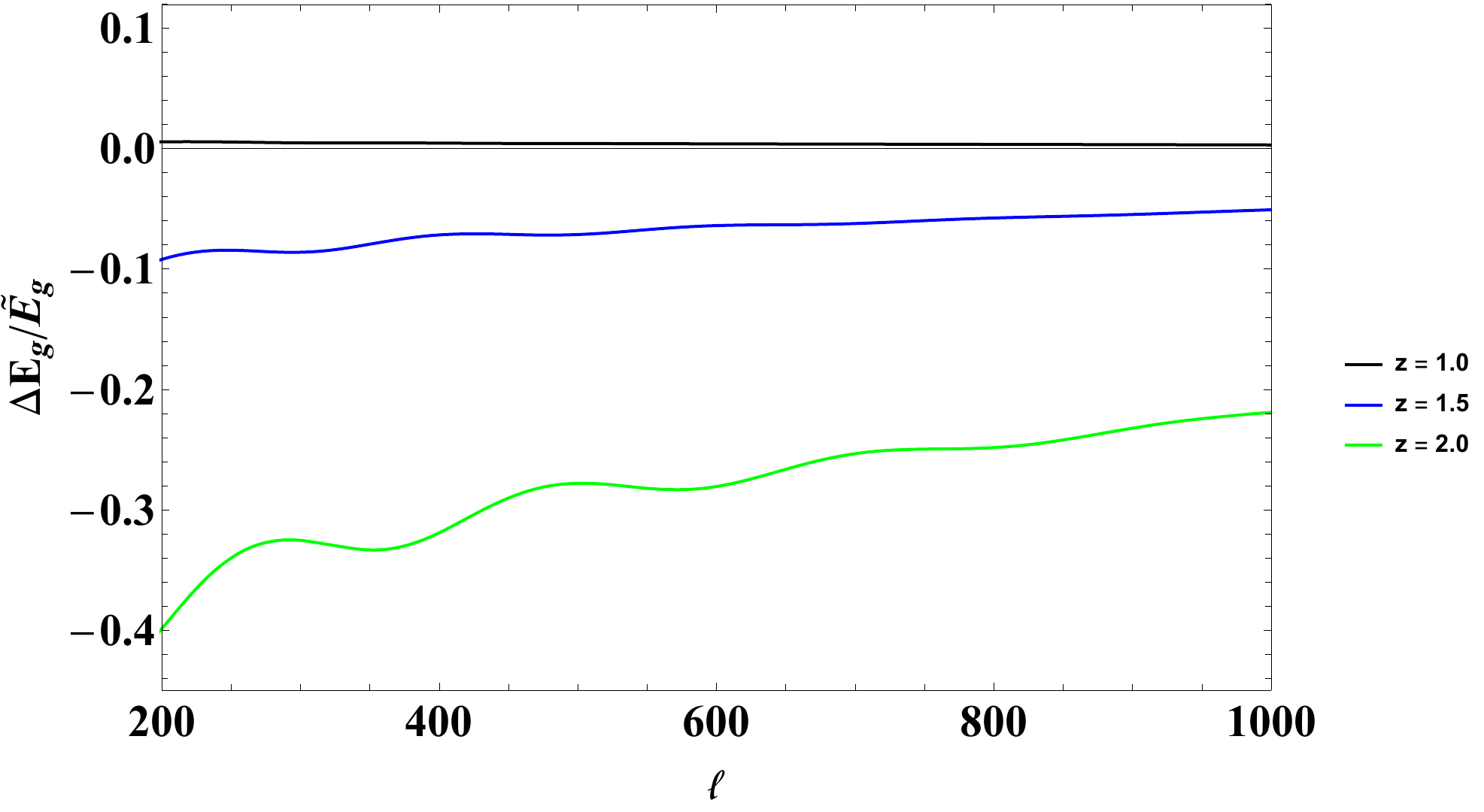}}
	\caption{Same as Fig.~\ref{f:des} but for the case of Euclid. At $z=1$ we have $s(z)=0.37$ which renders the correction terms very small, while at $z=2$ we find $s(z)=1.1$. The shaded area represents the cosmic variance error as a function of $\ell$ for $\ell>100$, here shown with respect to the highest redshift $z=2$..}
	\label{f:euclid}
\end{figure}

As is seen in fig.~\ref{f:euclid}, now the uncorrected ${E}_g$ has an error of about 40\% compared to the corrected $\tilde E_g$ for the highest redshift of Euclid, that is, $z=2$. The error is significantly scale dependent which comes from the fact that the uncorrected signal is scale dependent.  As one can see from the top panel, the corrected statistics, $\tilde E_g(\ell,z)$ is again scale independent for $\ell\gsim 50$. Also in this case, the constant behavior extends significantly beyond the naive non-linearity scale $\ell_L=k_L\chi(\bar z)$. A given value $\ell$ with corresponding wave number $k=\ell/\chi(\bar z)$ truly corresponds to a mean wavelength (taking into account an average radial contribution)
\be\label{e:wlen}
\bar\la(\ell, \bar z) \simeq \sqrt{(2\pi\chi(\bar z)/\ell)^2 +(\De z/2H(\bar z))^2}\,.
\ee
We find that for $\ell>300$ this mean wavelength is dominated by the second (radial) term for $z=2$ while this domination starts a bit earlier (roughly at $\ell=150$) for $z=1$. For this reason also for $\ell\sim 1000$ our result comes mainly from the linear regime and is expected to be scale independent.
The main effect of non-linearities is that they somewhat smooth out the acoustic oscillations.

\begin{figure}[H]
	%\vspace*{-1cm} ~~  \\
	\centering
	{\includegraphics[width=0.7\textwidth]{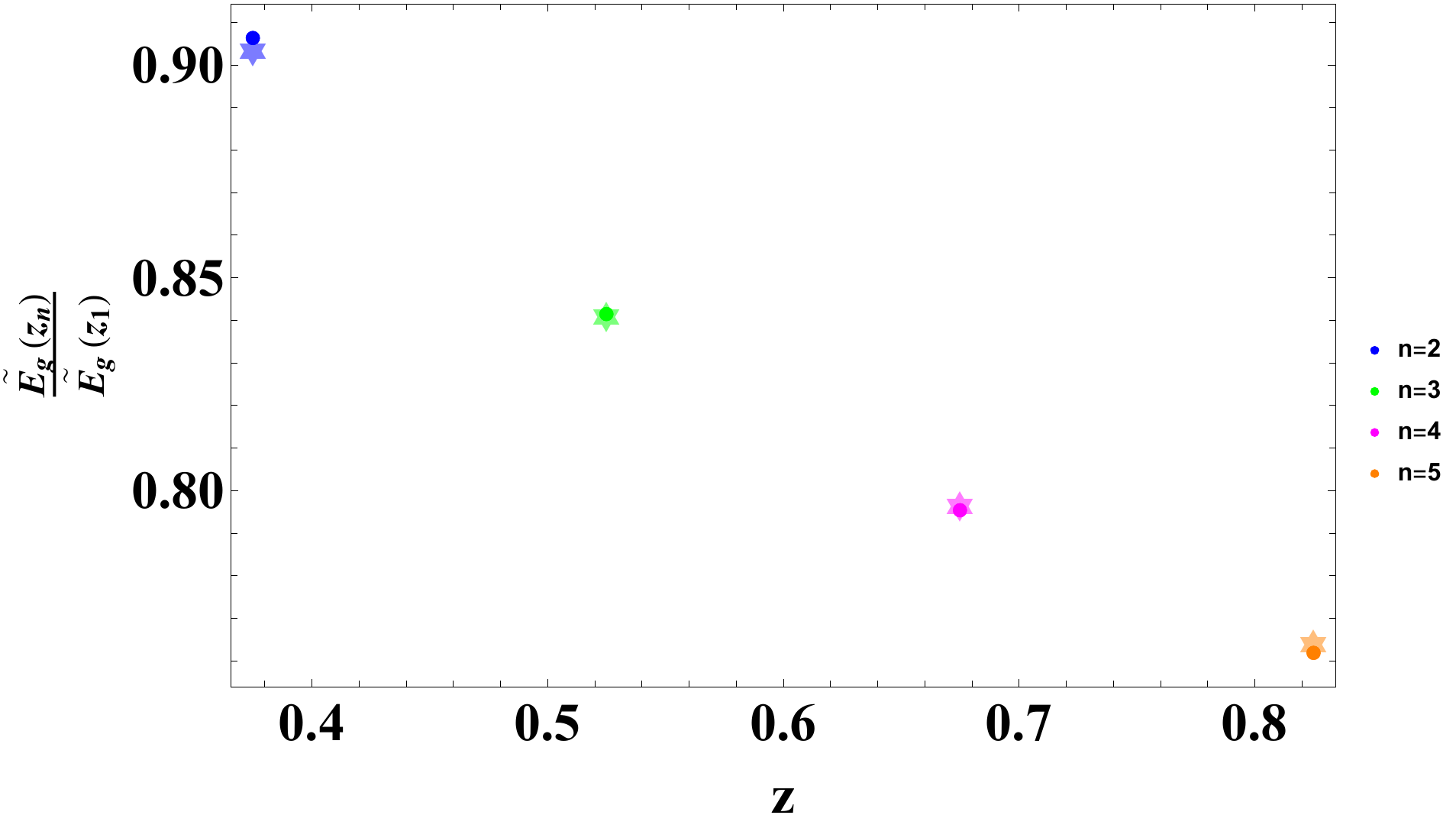}}
	\caption{The ratio as defined in eq.(\ref{e:ratio}), here shown for the DES redshift bins. The circular dots represent the theoretically expected values as in the right hand side of eq.~\eqref{e:ratio}, and the stars represent our numerical results. }
	\label{f:egratio}
\end{figure}

As for the absolute value of the result we have found that our normalization $\Ga(z)$ has to be multiplied by an overall factor which amounts to 6 for DES and 
0.35 for Euclid. The ratio,
\be
\frac{E_g(\ell,z_2)}{E_g(\ell,z_1)} = \left(\frac{\Om_m(z_1)}{\Om_m(z_2)}\right)^{0.55}
\label{e:ratio}
\ee
which we expect in a $\La$CDM cosmology
is very well realized in our findings (fig.~\ref{f:egratio}).

\section{Discussion and Conclusion}\label{s:con}
With upcoming surveys aiming for high redshift probes and increasing precision, the utility of the $E_g$ statistics in measuring deviations from general relativity is very promising. It is however important to include lensing and magnification bias effects in these measurements to reach the best possible accuracy. In this paper we have shown how this can be achieved in a photometric survey where we dispose of galaxy number counts, shear measurements and their correlation so that all three observable spectra $C_\ell^{gg}$, $C_\ell^{\ka g}$ and $C_\ell^{\ka\ka}$ are available. 
Our method constructs the corrected $E_g$ statistics termed $\tilde E_g(\ell,z)$  from these observables. In the linear regime, this quantity is independent of scale (for sufficiently large $\ell$) and bias. For redshift bins of the size considered here, the linear regime extends all the way up to $\ell=1000$ due to the relatively large radial contribution to the effective wavelength, see eq.~(\ref{e:wlen}).

We have found that  the lensing contributions are small for low redshift surveys like DES (up to $\sim 4\%$), but they are very significant in case of higher redshift surveys like Euclid (up to $\sim 40\%$).  Our work is based purely on theoretical calculations using linear  perturbation theory and halofit and has to be taken with a grain of salt.  From  the DES survey we simply assume the same foreground redshifts, galaxy bias and galaxy distribution in redshift, while for Euclid we use the galaxy distribution in redshift  as well as the galaxy and magnification biases. Nevertheless, it is safe to assume that Euclid will be able to measure the needed $C_\ell$'s with cosmic variance limited accuracy out to $\ell\sim 1000$. For $z=2$ this corresponds to a  transversal (minimal) scale of about 14$h^{-1}$Mpc which is still well in the linear regime at $z=2$.
The relative error from cosmic variance~\cite{Durrer:2008aa} of a $C_\ell$ observable is $\sqrt{2/(2\ell+1)}$ so that 
$\tilde E_g$ which is a ratio of  $C_\ell$'s is expected to have a cosmic variance error of
\be
\frac{\de^{cv}\tilde E_g(\ell)}{\tilde E_g(\ell)} =  \frac{2}{\sqrt{2\ell+1}}\simeq \sqrt{\frac{2}{\ell}} \,,
\ee
which amounts to 14\% for $\ell=100$ and 4.5\% for $\ell=1000$.  This is the cosmic variance indicated as shaded region in fig.~\ref{f:euclid}. When summing all available multipoles $\ell$ (assumed to be independent) from some value $\ell_{\min}$ to $\ell_{\max}$, the cosmic variance error can be further reduced to
\be
\frac{\de^{cv}\tilde E_g}{\tilde E_g} \simeq  \frac{\sqrt{2\log\left(\ell_{\max}/\ell_{\min}\right)}}{\ell_{\max}-\ell_{\min}}\,,
\label{e:cosvar}
\ee
which is about 0.003, significantly less than 1\%, for the values $\ell_{\max}=1000$ and $\ell_{\min}=100$ (see fig.~\ref{f:cosvar} below). 
\begin{figure}[H]
	%\vspace*{-1cm} ~~  \\
	\centering
	{\includegraphics[width=0.8\textwidth]{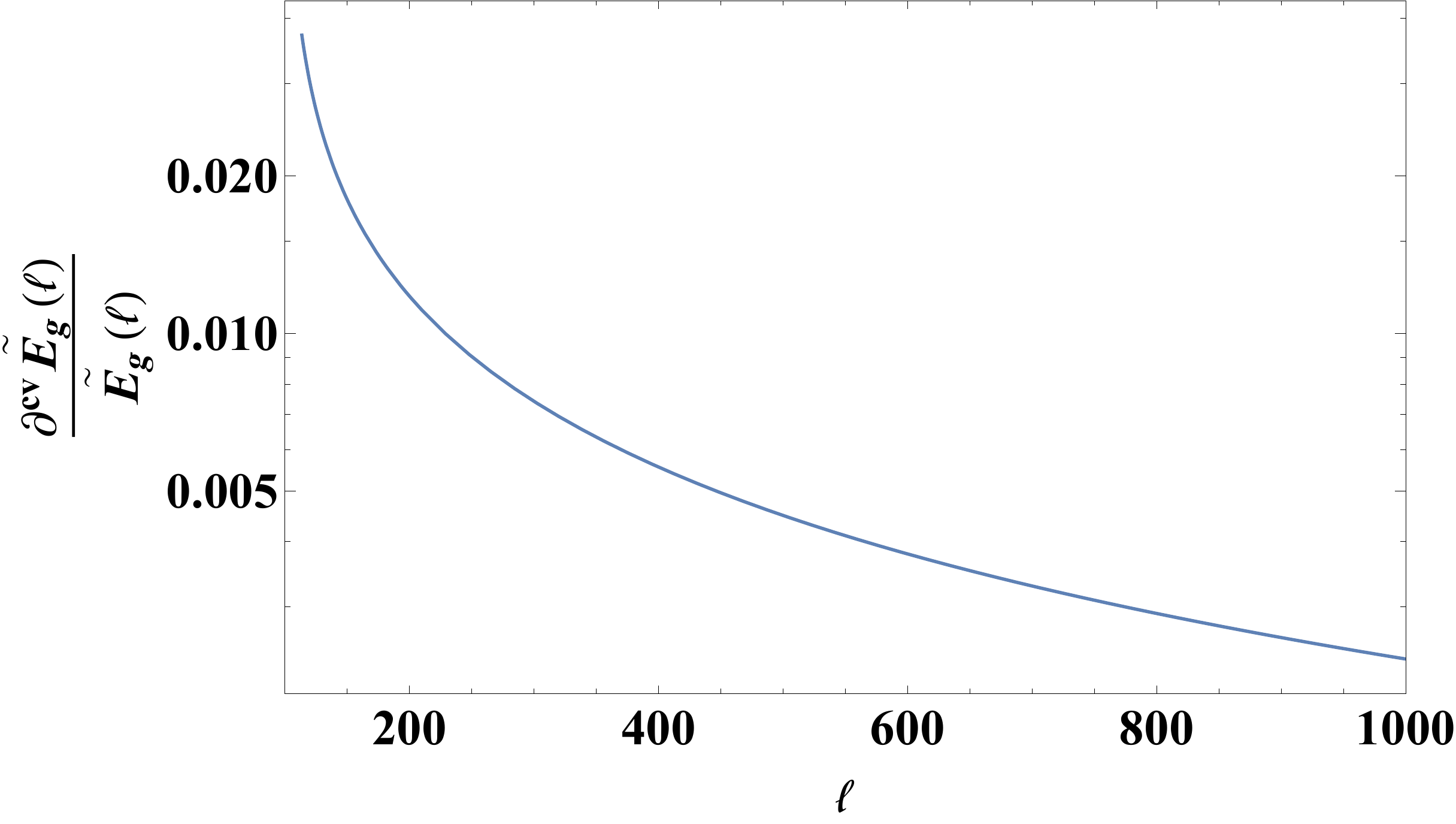}}
	\caption{The cumulative cosmic variance error as a function of $\ell$ as given in eq.(\ref{e:cosvar}).}
	\label{f:cosvar}
\end{figure}

This shows that the corrections discussed in this work, even if  probably not yet relevant for the present DES data, certainly have to be taken into account to optimize our analysis of future data of the quality of Euclid. The strength of the $E_g$ statistics is its scale invariance which allows a simple combination of the results from different scales. In addition, its bias independence renders it independent of the galaxies considered as long as their bias is linear, even if it is not scale independent. However, at second order we in general expect non-linear bias which will affect the $E_g$ statistics which renders it less attractive at small, non-linear scales.

\section*{Acknowledgement}
We thank Azadeh Moradinezhad Dizgah for helpful comments and suggestions on a first draft.
This work is supported by the Swiss National Science Foundation.

\appendix

\bibliography{refs}
\bibliographystyle{JHEP}
\end{document}